\documentstyle[aps,prd,psfig,floats,amssymb,colordvi]{revtex}

\def \be{\begin{equation}}
\def \ee{\end{equation}}
\def \bea{\begin{eqnarray}}
\def \eea{\end{eqnarray}}

\def \s2{\sqrt 2}

\begin{document}

\draft

\twocolumn[\hsize\textwidth\columnwidth\hsize\csname @twocolumnfalse\endcsname

\title{Scalar field induced oscillations of neutron stars and
gravitational collapse}

\author{
Florian Siebel$^{(1)}$, 
Jos\'e A. Font$^{(1)}$
and
Philippos Papadopoulos$^{(2)}$
}

\address{
$^{(1)}$Max-Planck-Institut f\"ur Astrophysik,
        Karl-Schwarzschild-Str. 1\\
        D-85741 Garching, Germany
\\
$^{(2)}$School of Computer Science and Mathematics,
        University of Portsmouth\\
        PO1 2EG, Portsmouth, United Kingdom
}

\date{\today}

\maketitle

\begin{abstract}
We study the interaction of massless scalar fields with
self-gravitating neutron stars by means of fully dynamic numerical
simulations of the Einstein-Klein-Gordon perfect fluid system. Our
investigation is restricted to spherical symmetry and the neutron stars are
approximated by relativistic polytropes. Studying the nonlinear
dynamics of isolated neutron stars is very effectively performed
within the characteristic formulation of general relativity, in which
the spacetime is foliated by a family of outgoing light cones. We are
able to compactify the entire spacetime on a computational grid and
simultaneously impose natural radiative boundary conditions and
extract accurate radiative signals. We study the transfer of energy
from the scalar field to the fluid star. We find, in particular, that
depending on the compactness of the neutron star model, the scalar
wave forces the neutron star either to oscillate in its radial modes
of pulsation or to undergo gravitational collapse to a black hole on a
dynamical timescale. The radiative signal, read off at future null
infinity, shows quasi-normal oscillations before the setting of a late
time power-law tail.
\end{abstract}


]

\section{Introduction}
\label{intro}

Obtaining reliable estimates for gravitational wave signals emitted
from the stellar collapse of massive stars is one of the key motivations for
numerical relativity. The strength of actual astrophysical sources is
still under investigation, as the relevance of such sources for the
first and second generation of interferometric detectors depends on
the details of signal amplitude and frequency, but also on the
occurrence rates (for a current view, see~\cite{fryer}). In any case,
relativistic collapse is a fundamental physical process, and the
development of relevant computational procedures has been a long
steady process over the past three decades 
(see, e.g.~\cite{stark,shibata,dimmelmeier}; see also~\cite{janka}
and references therein). In contrast to a Newtonian approximation, where 
the computational problem is well defined and attention can be devoted to
astrophysical details~\cite{janka,ewald}, there is no consensus as to what 
is the optimal, or at least adequate, framework for developing relativistic
simulations.

We will use here the so-called {\em characteristic} formulation of
general relativity~\cite{BBM62,Sac62}. The formalism has been
developed specifically for addressing ambiguities concerning
gravitational radiation and is well adapted to handle the propagation
of signals. It allows for spacetime compactification, which avoids
problems due to the artificial reflection of the fields at outer
boundaries.  In addition, it allows for the extraction of physically
relevant global quantities like the News function and the Bondi
mass. Nevertheless, the computation of the dynamics of {\em sources}
of signals is a separate issue altogether, which has not been
addressed in this early work. The framework for computing a complete
spacetime within the characteristic approach has been laid out 
in~\cite{TaW66} and more explicitly in~\cite{IWW}. The translation of this
framework into a computational tool for vacuum spacetimes, in any
dimensions, has been a largely successful process (see~\cite{WLiv}
and references therein).

There are some noteworthy issues. Firstly, the domain of applicability
is limited to configurations in which the set of light-cones that
forms the basis of the coordinate system does not fold itself into
caustics. This puts a limit, for example, on the type of binary
systems that can be studied. Secondly, to obtain a complete spacetime
one must include in the computational domain the vertex of the
light-cones. This involves regularity conditions (and for explicit
integration methods, severe time-step restrictions), which at present
have been resolved only up to axisymmetric configurations~\cite{GPW94}.
Nevertheless, the approach has remarkable economy and stability, which
makes it a good candidate for studies of {\em isolated} relativistic
objects emitting gravitational radiation. For the simulation of
realistic sources, one would have to include suitable matter
models. It was demonstrated in~\cite{PaF99} that the modern techniques
of high-resolution shock-capturing schemes for solving the equations
of relativistic fluid dynamics can be effectively integrated within
this framework. A separate study implementing the approach focused on
the gravitational radiation properties of an accreting black
hole~\cite{PaF01}.

This work is the first example of the use of characteristic numerical
relativity for the study of dynamical neutron star spacetimes,
collapse and radiative signals. The present numerical study is 
performed in spherical
symmetry and uses a self-gravitating, massless, scalar field.  The
scalar field serves as a simple matter model which mimics
gravitational waves. It has been used frequently to study global
properties of spacetimes, black hole formation and radiative
signals. For the latter, the work focused on the interaction of scalar
waves and black holes~(e.g.~\cite{GoW92,MaC96}) and especially on the
emergence of power-law tails~\cite{GPP94,GPP942,GWS94,PaL97}, which
arise from late time backscattering of the scalar field at the
exterior spacetime geometry~\cite{price}. There are few studies of the 
interaction of scalar fields with fluid stellar objects. 
Recently~\cite{HaC00} analyzed the scattering of scalar fields off boson 
stars and the emergence of critical solutions for this setup. They found 
that the scalar field can either make the boson star collapse to a black 
hole or to disperse its mass to infinity. 

Time-dependent simulations of scattering of gravitational waves
packets with neutron stars, as a means of computing the frequency
spectrum of the neutron star (see, e.g.~\cite{KoS99} for a
recent review), have been studied by Allen et
al~\cite{AAKS97} for polytropic equations of state (EoS), and by 
Ruoff~\cite{ruoff} for more realistic EoS. Such simulations were 
performed using linear perturbation codes. Pavlidou et al.~\cite{PTBS00}
studied the radiative falloff of scalar fields in neutron star spacetimes, 
using (idealized) analytic, constant density neutron star models and
assuming stationarity for the fluid and the geometry.

In the present work we investigate the nonlinear dynamics of neutron
stars interacting with scalar fields.  We are especially interested in
the following questions: How does a stable neutron star react when it
interacts with the scalar field? Can the scalar field force the star
to undergo gravitational collapse? What is the result of the interaction
on the scalar field? In order to answer these questions we perform 
spherically symmetric, coupled evolutions of the Einstein-Klein-Gordon
perfect fluid system. We study the reflection of finite scalar wave 
packets off neutron stars for a series of stars parametrized by the 
central density for a given polytropic EoS with polytropic index $n=1$.
Our study focuses on the fate of the system during the interaction,
both on the generation of nonlinear oscillations in the neutron star
and on the gravitational collapse of the neutron star to a black hole. 

The paper is organized as follows: Section~\ref{framework} describes the 
basic mathematical foundations of our approach. In Section~\ref{implementation}, 
we briefly discuss the numerical techniques and the implementation used in the
simulations. In Section~\ref{tests}, we present several numerical tests 
of our code, aimed to assess the correct implementation of its different 
components, the hydrodynamic evolution, the scalar field evolution and the
metric solver. Section~\ref{simulations} describes the actual numerical 
investigation on the interaction between the neutron stars and the scalar field.
Finally, Section~\ref{summary} summarizes our findings. Throughout the paper
we use geometrized units $G=c=1$ and further assume that $M_{\odot}=1$.
Greek indices run from 0 to 3.

\section{Mathematical Framework}
\label{framework}

We consider a general spherically symmetric spacetime with a two
component stress energy tensor of a perfect fluid and a scalar field,
$T^{\mu\nu} = T^{\mu\nu}_{F} + T^{\mu\nu}_{\Phi}$. The geometry of our
setup follows the lines of the
Tamburino-Winicour-formalism~\cite{TaW66}, in particular as it is
applied in regular spacetimes, where the foliation of light-cones emanates
from a freely falling central observer~\cite{IWW}.

\subsection{Einstein equations}
\label{metric}

By adopting the Bondi-Sachs~\cite{BBM62,Sac62} form of the metric element
in spherical symmetry,
\begin{equation}
\label{eq:bondi-sachs}
ds^2 = - \frac{e^{2\beta}V}{r} du^2
- 2 e^{2\beta} du dr + r^2 (d\theta^2 + \sin\theta^2 d\phi^2),
\end{equation}
the spacetime geometry is completely described by the two functions
$\beta(u,r)$ and $V(u,r)$.

A sufficient set of Einstein equations for obtaining the spacetime
development are grouped as
\begin{eqnarray}
\label{g1}
G_{ur} & = & \kappa T_{ur}, \\
\label{g2}
G_{rr} & = & \kappa T_{rr}, \\
\label{g3}
G_{uu}|_{\Gamma} & = & \kappa T_{uu}|_{\Gamma},
\end{eqnarray}
where the $u$ coordinate is defined by the level surfaces of a null
scalar (i.e., a scalar $u$ satisfying $\nabla^{\mu} u \nabla_{\mu} u =
0$). The $r$ coordinate is chosen to make the spheres of rotational
symmetry have area $4 \pi r^2$. The $x^{2},x^{3}$ coordinates in this
geometry are simply taken to be the angular coordinates
$(\theta,\phi)$ propagated along the generators of the null
hypersurface, i.e., they parameterize the different light rays on the
null cone. With our choice of units the constant $\kappa$ is simply
$\kappa = 8 \pi$. The first two Einstein equations, Eqs.~(\ref{g1})
and~(\ref{g2}), contain only radial derivatives
and are to be integrated along each null surface. The last
equation~(\ref{g3}) is a {\em conservation} condition, satisfied on
the vertex of the null cones $\Gamma$ due to the regularity 
conditions. We choose $\Gamma$ to be a timelike geodesic which 
coincides with the origin of a neutron star at $r=0$. 
Equation~(\ref{g2}) may be substituted for by the equivalent 
expression $g^{AB}R_{AB}=8\pi g^{AB}(T_{AB}-g_{AB}T/2)$, where 
$R_{\mu\nu}$ is the Ricci tensor and the indices $(A,B)$ run over the
angular coordinates $x^{2},x^{3}$.

Using the line element and Eqs.~(\ref{g1}) and~(\ref{g2})
the $\beta$ and $V$ hypersurface equations are given by
\begin{eqnarray}
\label{br}
\beta_{,r} & = & 2 \pi r T_{rr} \, ,\\
\label{Vr}
V_{,r}& = & e^{2\beta} (1 - 4 \pi r^2 (g^{AB}T_{AB} - T)) \, .
\end{eqnarray}
The comma in the above equations indicates, as usual, partial
differentiation. Boundary conditions for
$(\beta(u)_{\Gamma},V(u)_{\Gamma})$ needed for the radial integrations 
are provided by imposing regularity at the origin, where the coordinate
system is assumed to be a local Fermi system, leading to
\begin{eqnarray}
\label{regb}
\beta &=& O(r^{2}),
\\
\label{regV}
V &=& r + O(r^{3}).
\end{eqnarray}
By imposing such condition at the origin, the
lapse of coordinate time $du$ is related to the 
corresponding lapse of ``retarded time" $d \tau$ measured
by distant observers at $r \to \infty$ as
\begin{equation}
\label{obst}
d \tau = e^{2H} du, 
\end{equation}
where 
\begin{eqnarray}
H = \lim_{r \to \infty} \beta.
\end{eqnarray}

\subsection{Scalar Field equations}
\label{scalar}

The dynamics of a scalar field $\Phi$ is governed by the minimally coupled
Klein-Gordon equation in spherical symmetry,
\begin{equation}
    \nabla_{\mu}\nabla^{\mu} \Phi = 0,
\label{eq:klein}
\end{equation}
where $\nabla_{\mu}$ is the covariant derivative. This is the equation of
motion for the Lagrangian
\begin{equation}
      L = -{1\over 2}\nabla_{\mu} \Phi\nabla^{\mu} \Phi \,,
\label{eq:lagran}
\end{equation}
with a corresponding stress energy tensor given by
\begin{equation}
    T^{\mu\nu}_{\Phi}= \nabla^{\mu} \Phi \nabla^{\nu} \Phi + L g^{\mu\nu}~.
\label{eq:ein}
\end{equation}

Using a characteristic foliation the scalar wave equation,
Eq.~(\ref{eq:klein}), takes the form
\begin{equation}
     2 (r \Phi_{,u})_{,r} = \frac{1}{r} (r V \Phi_{,r} )_{,r}.
\label{eq:SWE}
\end{equation}
In terms of the intrinsic 2-metric of the $(u,r)$ sub-manifold,
\begin{equation}
    \eta_{CD} dx^{C} dx^{D}
    =-e^{2\beta}du({V \over r}du+2dr),
\end{equation}
where the indices $(C,D)$ run over the coordinates $(u,r)$, 
Eq.~(\ref{eq:SWE}) reduces to
\begin{equation}
     \Box ^{(2)}g= {e^{-2\beta} g\over r} \left({V\over r}\right)_{,r},
\label{eq:hatwave}
\end{equation}
where  $g=r\Phi$ and $\Box ^{(2)}$ is the D'Alembertian operator
associated with ${\eta}_{CD}$.

\subsection{Hydrodynamic equations}
\label{hydro}

The evolution of the fluid is determined by the local conservation laws
of stress energy and density current
\begin{eqnarray}
\label{eq:conT}
\nabla_{\mu} T^{\mu\nu}_{F} & = & 0, \\
\label{eq:conti}
\nabla_{\mu} (\rho u^{\mu}) & = &0, 
\end{eqnarray}
where $T^{\mu\nu}_{F}$ is the stress energy tensor of a perfect fluid
\begin{equation}
T^{\mu\nu}_{F} = \rho h u^{\mu} u^{\nu} + p g^{\mu\nu}.
\end{equation}
All quantities in the above expression have their usual meanings:
$\rho$ is the mass density, $h= 1 + \varepsilon+p/\rho$ is the specific
enthalpy, $\varepsilon$ is the specific internal energy and $p$ is the
pressure of the fluid. Moreover, $u^{\mu}$ is the four-velocity which
satisfies the normalization condition $g_{\mu\nu} u^{\mu} u^{\nu} = -1$.

Following~\cite{PaF99}, after introducing the definitions
$D = \rho u^{0}$, $S^{r} = T^{0r}_{F}$ and $E = T^{00}_{F}$, the fluid
equations can be cast into a first-order flux-conservative, hyperbolic system 
for the state-vector
${\bf U}=(D,S^r,E)$ :
\begin{eqnarray}
\label{eq:Dr}
D_{,u} + F^{r0}_{,r} & = &
- (\ln{\sqrt{-g}})_{,u} D   
\nonumber \\ && -(\ln{\sqrt{-g}})_{,r} F^{r0} \, ,\\
\label{eq:Sr}
S^{r}_{,u} + F^{r1}_{,r} & = &
- (\ln{\sqrt{-g}})_{,u} S^{r} 
\nonumber \\ & &-  (\ln{\sqrt{-g}})_{,r} F^{r1}
- \Gamma^{r}_{\mu\nu} T^{\mu\nu}_{F} \, ,\\
\label{eq:Er}
E_{,u} + F^{r4}_{,r} & = &
- (\ln{\sqrt{-g}})_{,u} E 
\nonumber \\ & &-  (\ln{\sqrt{-g}})_{,r} F^{r4}
- \Gamma^{u}_{\mu\nu} T^{\mu\nu}_{F} \, ,
\end{eqnarray}
where $\sqrt{-g}=r^{2}\sin\theta e^{2\beta}$ is the four
dimensional volume element and $\Gamma^{\alpha}_{\mu\nu}$ are the
Christoffel symbols. The precise form of the vector of fluxes
${\bf F}$ can be obtained by using Eqs.~(\ref{eq:conT})-(\ref{eq:conti})
(see also~\cite{PaF99}). The explicit relations between the primitive
variables ${\bf w}=(\rho, \varepsilon, u^{r})$ and the conserved variables
${\bf U}=(D,S^{r},E)$, for a perfect fluid EoS,
$p=(\Gamma-1)\rho\varepsilon$, where $\Gamma$ is the adiabatic index
of the fluid, are given in~\cite{PaF99}.

With the above definitions, the metric equations~(\ref{br})-(\ref{Vr})
read, for the combined stress energy tensor of a fluid-scalar field
system,
\begin{eqnarray}
\label{eq:beta} \beta_{,r} & = & 2 \pi r  (\rho h (u_r)^2 +
(\Phi_r)^2) \, ,
\label{metric1}
\\ \label{eq:V} V_{,r}& = & e^{2\beta} (1 - 4 \pi
r^2 (\rho h - 2 p)) \, .
\label{metric2}
\end{eqnarray}
Following~\cite{LFJMP01} we express the hydrodynamic quantities on the
right-hand side of Eqs.~(\ref{metric1})-(\ref{metric2}) solely in terms
of the conserved hydrodynamic quantities ${\bf U}$. This
avoids additional iterations when using explicit algorithms to
solve these ordinary differential equations.

In summary, the initial value problem consists of
equations~(\ref{g3}), (\ref{eq:hatwave}),
(\ref{eq:Dr}-\ref{eq:Er}), (\ref{eq:beta},\ref{eq:V}), the
scalar field initial data $\Phi(r,u_0)$ and initial and boundary data for
the fluid variables $(\rho,\varepsilon,u^{r})$ on the initial slice
$\Sigma_0$ (at time $u_0$). Those equations and initial data are
sufficient for obtaining a global solution to the problem.

\subsection{Global quantities}
\label{global}

Making use of the characteristic formulation of general relativity and
covering the infinite range of the radial coordinate with a finite
grid allows us to refer to some global quantities of the spacetime
such as the Bondi mass and the news function. Apart
from their physical relevance, these quantities can be used in
global tests of our numerical evolutions, as we will show below.

Instead of extracting the Bondi mass directly at future null infinity
we use the expression 
\begin{equation}
M = 4 \pi \int_{0}^{\infty} r^{2} e^{-2 \beta} T_{ru} dr
\end{equation}
for the Bondi mass at time $u$ in our numerical implementation.
Similarly, the news can be rewritten as~\cite{GoW92}
\begin{equation}
\label{NNews}
N = \frac{1}{2} e^{-2 H} \int_{0}^{\infty} \frac{V}{r} \Phi_{,r} dr.
\end{equation}
With these definitions, global energy conservation can be established,
\begin{equation}
\label{globcon}
M(u)-M(0) = \int_{0}^{u} - 4 \pi N(\hat{u})^{2} e^{2 H(\hat{u})} d\hat{u}.
\end{equation}

\section{Numerical Implementation}
\label{implementation}

In order to study the interaction of the scalar field and the neutron star
in a global spacetime we use nonequidistant grids for the radial
coordinate $r$. Furthermore, to avoid dealing with complicated
stencils in the numerical implementation, we make use of the following
procedure, generalizing previous implementations in characteristic numerical
relativity (see, e.g.~\cite{GPW94}): Starting with an equidistant grid
in the coordinate $x \in \left[0,1\right]$, we allow for a general coordinate
transformation $r = r(x)$. Using the chain rule we rewrite the partial
derivatives appearing in the above equation with
\begin{equation}
(\ ),_{r} = (\ )_{,x} \ \frac{dx}{dr},
\end{equation}
thus effectively rewriting all our
equations in the coordinate $x$. Unless otherwise stated we use the
relation
\begin{equation}
\label{radial}
r = \frac{15x}{1-x^4}
\end{equation}
for all computations presented in this work. Using such a coordinate
transformation the repartition of grid points in the coordinate $r$ is
almost equidistant for small radii and gets infinitely sparse for $x \to 1$,
which corresponds to future null infinity ${\cal{J}}^{+}$.

We use a second order Runge Kutta method to solve the metric 
equations~(\ref{metric1}) and~(\ref{metric2}). To determine 
the equilibrium models of
our relativistic stars, we also use the Runge Kutta method to solve
the so-called Tolman-Oppenheimer-Volkoff equations, formulated on a
null hypersurface as in~\cite{PaF99}.  
 
The integration of the evolution equation for the scalar field,
Eq.~(\ref{eq:hatwave}) (or equivalently Eq.~(\ref{eq:SWE})), proceeds
with the specification of initial data $\Phi(u_0,r)$ on the initial
null cone $u_{0}$. For the characteristic evolution we have used two
different algorithms, both marching from the origin to the exterior.

The first procedure is based upon the construction of a null
parallelogram built up from incoming and outgoing radial
characteristics\cite{GWI92}. In this procedure one needs first to 
determine the right hand side of Eq.~(\ref{eq:hatwave}) at the
center of the parallelogram to the desired order of accuracy. Then,
an integral relation between this source term and the values of $g$
at the four corners of the parallelogram -
which do not necessary have to coincide with
grid points - has to be employed in order to
compute the scalar field at that corner of the parallelogram lying
next to the grid point which is to be updated. Suitable interpolations
then give the scalar field at the new grid point to second order
accuracy.

The second alternative procedure we have implemented to
solve the scalar field equation is based on a direct discretization of
Eq.~(\ref{eq:SWE}) using a second order, finite difference,
nondissipative algorithm discussed in~\cite{Leh00}.

Due to the stencils of both algorithms, we cannot use them at the
origin, where a regular behavior of the scalar field as
$\Phi = a + b r + c r^{2}$ is assumed. The linear term introduces a
kink at the origin, but it is necessary in our foliation - as can be seen 
from the analytic solution for the wave equation in Minkowski space 
consisting of an ingoing and outgoing wave. Note that the scalar field
enters the metric only through Eq.~(\ref{eq:beta}), thus respecting the 
regularity conditions given by Eqs.~(\ref{regb},\ref{regV}). 
Substituting this ansatz into
Eq.~(\ref{eq:SWE}) and grouping those terms with the same powers of
$r$ we find that $a_{,u} = b$, $b_{,u} = 1.5 c$. Extracting the
coefficients $a,b$ and $c$ on the null cone $u_{0}$, we update $a$ and
$b$ to obtain the scalar field at the first two
grid points of the new hypersurface, which then allows us to start
the marching procedure along the null hypersurface with either of
the two algorithms described above.

By experimenting with both algorithms, we found that, on the one
hand, the scheme based upon a direct discretization of the wave
equation is more accurate in
the long-term behavior in the interior of the numerical domain.
This was relevant to resolve the late time fall-off behavior
of the scalar field, as we describe below in Section~\ref{simulations}.
On the other hand, the algorithm based upon the null parallelogram
is, however, superior close to future null infinity, where we
regularized the equations following the work of~\cite{GPW94}. Therefore,
for the results presented in this work we have used a ``hybrid
algorithm'', in which a direct discretization of Eq.~(\ref{eq:SWE})
is used in the interior of the computational domain and the parallelogram
algorithm is used close to future null infinity $\cal{J}^{+}$.

Concerning the numerical integration of the system of hydrodynamic
equations, its hyperbolic mathematical character allows 
for a solution procedure based on the computation of (local) Riemann
problems at each cell-interface of the numerical grid. At cell $i$
the state-vector ${\bf U}$ is updated in time (from $u^n$ to $u^{n+1}$) 
using a conservative algorithm
\begin{eqnarray}
    {\mathbf U}_{i}^{n+1}={\mathbf U}_{j}^{n}-\frac{\Delta u}{\Delta x}
    (\widehat{{\mathbf F}}_{i+1/2}-\widehat{{\mathbf F}}_{i-1/2}) +
    \Delta u {\mathbf S}_j \,,
  \end{eqnarray}
\noindent
where the numerical fluxes, $\widehat{{\mathbf F}}$, are 
evaluated at the cell interfaces according to some particular
{\it flux-formula} which makes explicit use of the full spectral
decomposition of the system. For our particular formulation of the
hydrodynamic equations such characteristic information was 
presented in~\cite{PaF99}.

In more precise terms the hydrodynamics solver of our code uses a 
second order Godunov-type algorithm, based on piecewise linear 
reconstruction procedures at each cell-interface~\cite{vanleer} and
the HLLE approximate Riemann solver~\cite{harten,einfeldt}.
General information on such schemes in relativistic hydrodynamics
can be found, e.g. in~\cite{fontlr} and references therein.

\begin{figure}[t]
\centerline{\psfig{file=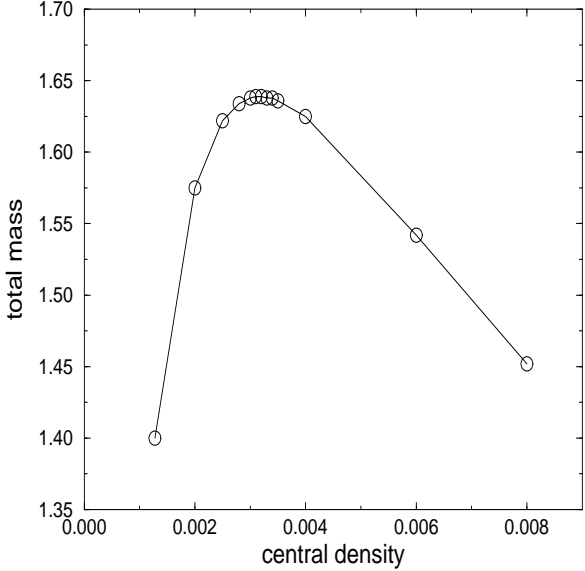,height=3.5in,width=3.5in}}
\caption{
Stability curve for neutron star models with the polytropic equation of state 
$p = K \rho^{\Gamma}$, with $K=100$ and $\Gamma=2$. Models lying to the left of 
the maximum of the curve, at about $\rho_{c} = 3.2 \times 10^{-3}$, are
stable against gravitational collapse. The circles are calculated with
our initial data solver and are connected by straight lines. We use units in
which $G=c=M_{\odot}=1$.}
\label{stab}
\end{figure}

\section{Code tests}
\label{tests}

We now present representative results obtained in the process
of code calibration. The assessment of the numerical implementation
is provided by comparing to previous results and by checking
global energy conservation tests.

\subsection{Null cone evolutions of self-gravitating, stable neutron stars}
\label{test:ns}

As a first step to validate our numerical code we start by studying its
ability to keep the equilibrium of initially stable neutron star models. For 
this purpose we perform long-term simulations of such initial data and analyze 
the stability of the code. Furthermore, we use these evolutions to compute the 
frequencies of the radial modes of pulsation. We compare the frequencies 
obtained with our nonlinear code to results of linear evolutions from 
perturbation theory. 

\begin{figure}[t]
\centerline{\psfig{file=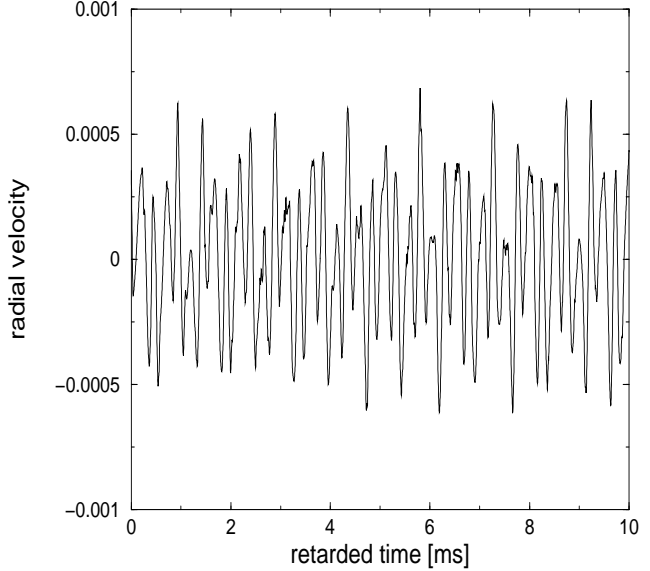,height=3.5in,width=3.5in}}
\caption{
Time evolution of the radial velocity $u^{x}$ at half radius of the star.
The neutron star model has a central density $\rho_{c} = 1.5 \times 10^{-3}$.
The oscillations are essentially undamped for the evolution shown, which
reflects the small viscosity of the hydrodynamic schemes employed.}
\label{freq1.5}
\end{figure}

\begin{figure}[h]
\centerline{\psfig{file=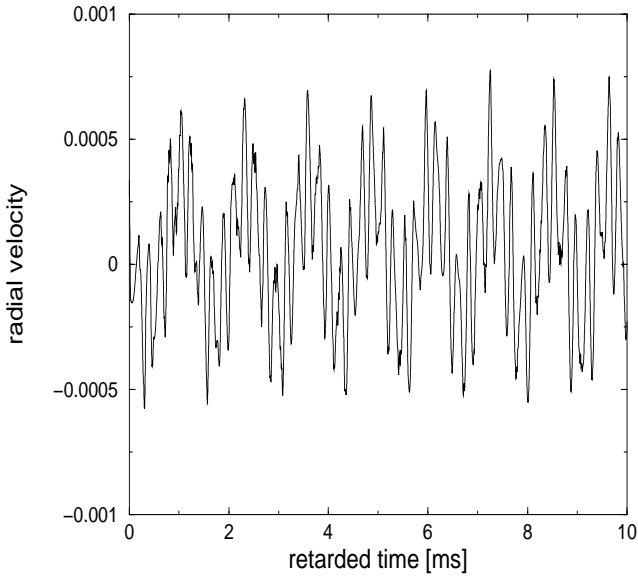,height=3.5in,width=3.5in}}
\caption{
Time evolution of the radial velocity $u^{x}$ at half radius of the star.
The neutron star model has a central density $\rho_{c} = 2.8 \times 10^{-3}$.}
\label{freq2.8}
\end{figure}

For all simulations presented in this paper the neutron star models are
approximated by a polytropic EoS, defined by $p=K\rho^{\Gamma}$, with
polytropic constant $K=100$ and adiabatic exponent $\Gamma\equiv 1+1/n =2$. 
Hence, the index of the polytrope is $n=1$. For 
the simulations presented in this section we choose two different models
with central density $\rho_{c}=1.5 \times 10^{-3}$ and $\rho_{c}=2.8 \times
10^{-3}$ (recall that we are using units in which $M_{\odot}=1$). Both 
models are located in the stable branch of the central density
- total mass - diagram (see Fig.~\ref{stab}).

When evolving these neutron star models with our numerical code we 
are able to maintain them in stable equilibrium for thousands of 
light-crossing times of the star without any sign of numerical 
instabilities.

To validate the code further we computed the frequencies of the radial 
modes of pulsation. For this aim we have to allow the star to
(radially) contract and  expand during the evolution. To this end, 
following~\cite{FMST98} 
(see also~\cite{FSK00}), we surround the star with a few zones representing
an artificial ``atmosphere" filling an otherwise vacuum region. The density 
in this atmosphere is set to sufficiently small values such that its presence 
does not affect the dynamics of the system. The typical values we choose 
are $10^{-7}-\ 10^{-8}$ times the central density of the star. Furthermore, 
to avoid any numerical problems due to (shock) heating in the atmosphere 
(the fluid in those zones is not in equilibrium and, therefore, it will 
collapse/accrete onto the neutron star), we follow the recipe described
in~\cite{FMST98} and enforce adiabatic evolution (by using the polytropic
EoS) in the atmosphere and in 
the outer layers of the neutron star (comprising the outermost 10 grid points).
After each time step, if the density has fallen below 1.5 times the
density of the atmosphere, the hydrodynamic quantities are reset to their 
atmosphere values. The innermost location where this procedure is done 
defines the radius of the star. As described in more detail in the next 
section the above values of the atmosphere density are small enough to 
guarantee conservation of energy despite the artificial resetting procedure.

\begin{figure}[t]
\centerline{\psfig{file=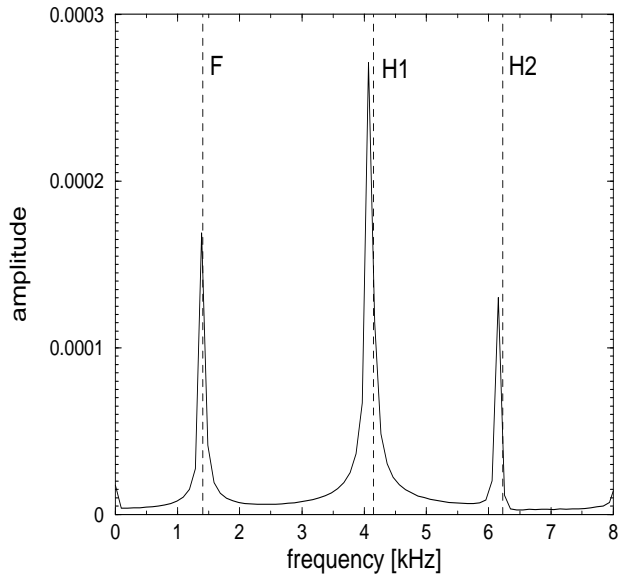,height=3.5in,width=3.5in}}
\caption{
Fourier transform of the time evolution shown in Fig.~\ref{freq1.5}.
The peaks in the Fourier transform indicate the mode frequencies of the
fundamental radial mode (around $f=1.4$ kHz) and the first two harmonics.
The neutron star model has a central density $\rho_{c} = 1.5 \times 10^{-3}$.
The dashed vertical lines indicate the corresponding frequencies obtained
with a perturbative (linear) code. The units in the $y$-axis are
arbitrary.}
\label{freq1.5m}
\end{figure}

\begin{figure}[h]
\centerline{\psfig{file=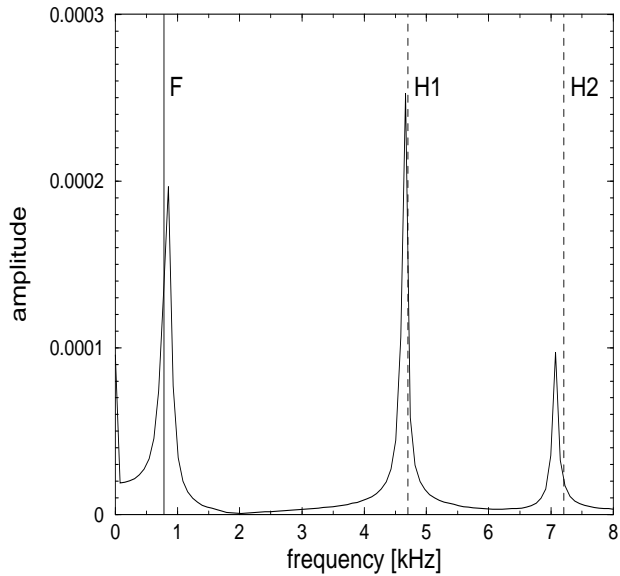,height=3.5in,width=3.5in}}
\caption{
Fourier transform of the time evolution shown in Fig.~\ref{freq2.8}.
The frequencies of the fundamental radial mode (around $f=0.8$ kHz) and the
first two overtones are shown for a neutron star with central
density $\rho_{c} = 2.8 \times 10^{-3}$.}
\label{freq2.8m}
\end{figure}

When evolving our stellar models in time, we find small deviations
around the equilibrium values due to the discretization errors. As a
result, the stars oscillate in their fundamental radial
modes. Figures~\ref{freq1.5} and~\ref{freq2.8} show the radial
velocity at half stellar radius for the above models as a function of
the retarded time measured by distant observers. These simulations
were performed with a grid of 800 zones covering the complete radial
domain. This amounts in using about half of the grid in resolving the
neutron star (We choose this resolution here to allow for comparisons
with the results of section V, where we have to resolve the scalar
field as well.)  As shown in~\cite{FSK00} one can use such evolutions
to obtain the frequencies of the excited modes of pulsation of the
star by simply Fourier transforming those data. In general, however, 
the excitation of the different modes by the truncation error of the
numerical schemes may not be sufficient to accurately determine the
mode frequencies.  Therefore, in order to compare unambiguously our
mode frequencies with perturbation theory, we further perturb the
density of the equilibrium models with an explicit eigenfunction $\rho
= \rho_{o} + A \rho_{c} \sin(\pi r/R)$, where $R$ denotes the radius 
of the star (see Table I) and $\rho_o$ is the density of the unperturbed 
star. The typical amplitude we use in the perturbation is $A=10^{-6}$.

Figures~\ref{freq1.5m} and~\ref{freq2.8m} show the frequencies of the
fundamental mode and the first two overtones obtained by a Fourier
transform of the radial velocity evolutions. The dashed vertical lines
in these plots were obtained using a linear code~\cite{Ruo} which
evolves in time the perturbation equations. The agreement between the
two codes is remarkable.  The fundamental mode of the model with
$\rho_{c} = 2.8 \times 10^{-3}$ is already rather small, the star
being close to the unstable branch. We note that the code is able to
obtain a much higher number of overtones. Nevertheless, for the sake
of clarity in the comparison and for our purpose of assessing the
correct numerical implementation, it is sufficient to show
only the first two harmonics.

\subsection{Scalar field dynamics in a regular spacetime}
\label{test:sf}

\begin{figure}[t]
\centerline{\psfig{file=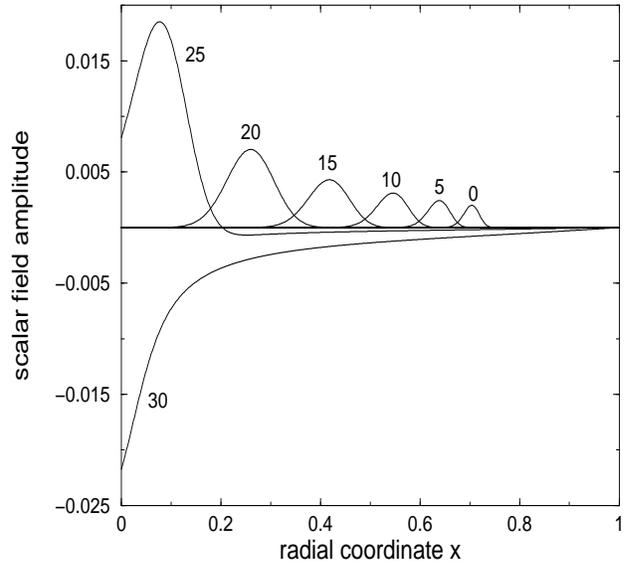,height=3.5in,width=3.5in}}
\caption{Radial profiles of the scattering of a scalar field off the
origin of the coordinate system at different times
of its evolution. The numbers in the plot refer to the time 
coordinate $u$. For times $u \ge 35$ the pulse has completely radiated
away leaving Minkowski spacetime behind.}
\label{phi}
\end{figure}

In this section we present results aimed to validate the numerical
implementation of the Einstein-Klein-Gordon solver. For this purpose
we investigate the reflection of a scalar field at the origin of the
coordinate system, turning off the hydrodynamics module of the code. 
The initial data for the scalar field packet are
\begin{equation}
\label{initial}
\Phi_{0} = 2 \times 10^{-3} e^{-(r-14)^{2}}.
\end{equation}
The location of this Gaussian pulse is chosen in such a way that, if
superposed on the neutron star spacetimes of the previous section, the
scalar wave pulse would initially lie outside the neutron star, the
overlap being completely negligible.
Evolving this data, the initial pulse approaches the origin, is
reflected, and radiates away, leaving behind Minkowski space. Such a
sequence can be followed in Fig.~\ref{phi}, for a simulation
employing a grid of 800
zones. We note the stability and smoothness of the solution, both at
the origin and at ${\cal J}^+$.

\begin{figure}[h]
\centerline{\psfig{file=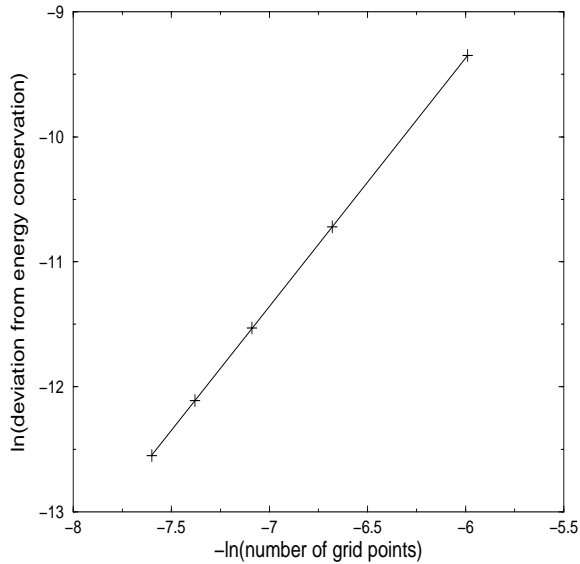,height=3.0in,width=3.0in,angle=-90}}
\caption{
Convergence test of the global energy conservation for the
Einstein-Klein-Gordon system once the scalar wave has completely 
radiated away after reflecting off the origin. The best linear fit 
to our results has a slope of 1.99, confirming the (global) second 
order accuracy of our numerical implementation.}
\label{converg}
\end{figure}

By evaluating global energy conservation, according to Eq.~(\ref{globcon}),
after the pulse has reflected off the origin, we find that the energy is 
conserved (as expected) to second order accuracy. The best linear fit to the 
curve shown in Fig.~\ref{converg} gives a slope of 1.99.

As an aside we note that by simply
changing the origin treatment in the code, it is possible to study
the evolution of a scalar field outside a spherical black hole. We 
performed such a simulation finding agreement with 
the results of~\cite{GWS94}.  

\subsection{Scalar field in a dynamic spacetime with a neutron star}
\label{sfns:test}

We consider now the full set of equations and prescribe initial data
consisting of the ingoing scalar field pulse given by
Eq.~(\ref{initial}), together with a stable equilibrium,
self-gravitating neutron star model with initial central density
$\rho_{c}=1.28 \times 10^{-3}$, $K=100$ and $\Gamma=2$. This
relativistic star model has a total mass of $1.4 \ M_{\odot}$ (see
also~\cite{FSK00}). We perform simulations of the scalar field
scattering off the neutron star, focusing our study in this Section on
the assessment of the global energy conservation properties of our
complete numerical implementation. A comprehensive study of the
dynamics of the scattering is deferred to Section~\ref{simulations}.

\begin{figure}[t]
\centerline{\psfig{file=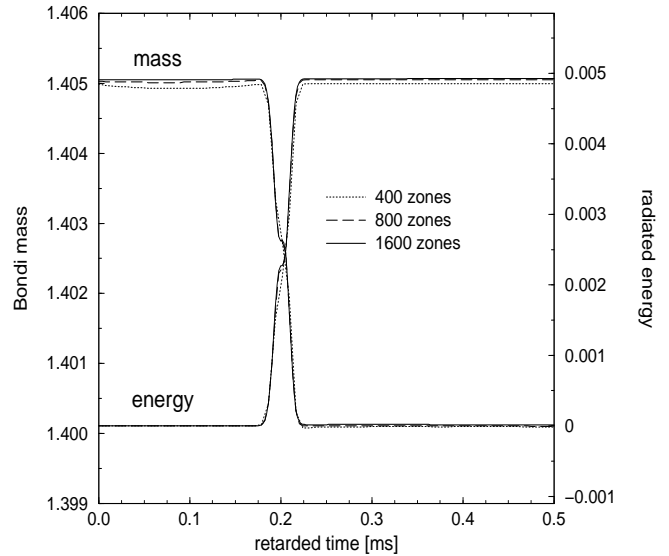,height=3.0in,width=3.2in}}
\caption{Bondi mass of the spacetime and total radiated energy of the
scalar field as a function of retarded time for different
resolutions.  At the beginning, the scalar field
contributes to the Bondi mass of the spacetime 
($1.405 M_{\odot}$),  before the Bondi mass drops in a small
time interval, when the main part of scalar field mass (about $5
\times 10^{-3} M_{\odot}$) is emitted to future null infinity ${\cal{J}}^{+}$. 
As the sum of the two curves is constant, the energy is globally
conserved (with a relative error of about $5 \times 10^{-6}$ for the
run with 1600 zones.)}
\label{mass}
\end{figure}

Fig.~\ref{mass} shows the Bondi mass of the neutron star - 
scalar field spacetime as a function of retarded time, combined with
the total mass of the scalar field radiated away to null infinity. 
As one can clearly see from this figure, the spacetime is losing mass 
exactly at the rate which is radiated to null infinity by the scalar field.

\begin{figure}[t]
\centerline{\psfig{file=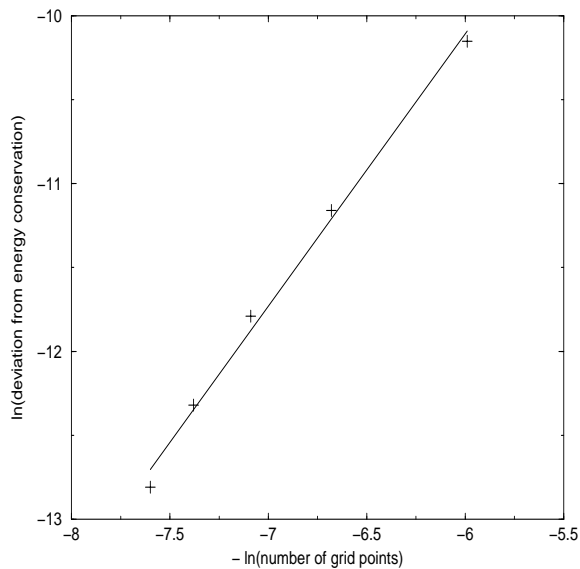,height=3.0in,width=3.0in,angle=-90}}
\caption{
Convergence test of the global energy conservation of a dynamic spacetime
containing a self-gravitating neutron star and a scalar field. The rate of 
convergence is 1.62. See text for details.}
\label{energy}
\end{figure}

By computing Eq.~(\ref{globcon}) at a fixed retarded time of
$\tau=0.5$ms for different grid resolutions, we find that our code conserves 
globally the energy with a convergence rate which lies in between 1 and
2. The best linear fit is depicted in Fig.~\ref{energy}. The fact that the 
convergence rate drops now below second order is, however, to be expected, 
since the approximate Riemann solver we are using for the integration of 
the hydrodynamic equations is only (locally) first order accurate at 
discontinuities (i.e., the surface of the star) and at local extrema (i.e., 
the center of the star) (see the related discussion in~\cite{FSK00}). 
Nevertheless, for the highest resolution we have used, 2000 radial 
grid points, the relative error in the energy conservation is of the 
order of $2 \times 10^{-6}$ for this very dynamical simulation.

\section{Dynamics of scalar field - neutron star interactions}
\label{simulations}

In this section we present our main results concerning the scattering of a
scalar field pulse off relativistic neutron stars. As mentioned before, we
use $n=1$ relativistic polytropes as neutron star models. All models we 
construct lie on the stable branch of the total mass-central density
diagram (see Fig.~{\ref{stab}) and are 
characterized by increasing central densities and compactness. Their 
basic properties are summarized in Table~\ref{nsmodels}.

\begin{table}
\caption{ 
Equilibrium properties of the $K=100$, $n=1$ neutron star models in
units in which $c=G=M_{\odot}=1$. The entries are as follows: $\rho_{c}$ 
is the central density, $M$ and $R$ are the mass and radius of the star, 
respectively, and $C=2M/R$ is the compactness parameter.}
\medskip
\begin{tabular}{cccc}
$\rho_{c} \,\, (10^{-3})$ & $M$ & $R$ & $C=2M/R$ \\
\hline
1.5  &  1.47  &  9.26  &  0.317
\\
2.2  &  1.60  &  8.45  &  0.379
\\
2.8  &  1.63  &  7.91  &  0.412
\\
2.9  &  1.64  &  7.84  &  0.418
\\
3.0  &  1.64  &  7.76  &  0.423
\\
\end{tabular}
\label{nsmodels}
\end{table}

In addition to the neutron stars we construct a Gaussian pulse of a
scalar field according to Eq.~(\ref{initial}), thus fixing the pulse
amplitude, width and location. For these initial data there is no
overlap between the star and the scalar field pulse which then makes it
possible to associate a specific initial mass with each one of the matter
fields.

In our setup we keep the central density as the only free parameter
of the simulations, choosing a unique polytropic EoS and fixing the
geometry and amplitude of the scalar field. This is clearly a severe 
restriction in the parameter space of the scattering problem. Nevertheless, 
we choose this particular setup since we are interested in investigating 
the relativistic effects of the interaction, where the scalar field has a 
strong impact on the dynamics of the neutron star. A detailed analysis of 
the whole parameter space is beyond the scope of this work.

When evolving in time the initial data, the scalar field travels
inwards, enters the neutron star and it is finally reflected at the
origin of the coordinate system. Contrary to the Einstein-Klein-Gordon
system without the perfect fluid,  which was discussed in 
Section~\ref{test:sf}, the presence of the neutron star,
and its associated potential well, may give rise to a phase of multiple
interactions of the wave back and forth the origin and the maximum of the
curvature potential. This, in turn, reflects itself in the existence of
quasi-periodic signals (trapped modes), as discussed by~\cite{PTBS00}
before its energy is radiated away. Furthermore, our neutron star models
have been chosen conveniently close to the maximum of the stability curve
(see Fig.~\ref{stab}). Depending on the compactness of the neutron 
star onto which the wave
pulse impacts, the stars are forced to either oscillate violently,
or to collapse to a black hole on a dynamical timescale. Fig.~\ref{spacetime} 
shows the spacetime diagram for the least compact neutron star model of our
sample, with $\rho_{c} =  1.5 \times 10^{-3}$. For this model, the scalar 
field is able to force the star to contract and to expand, pulsating 
radially, as can 
clearly be identified in the varying location of the star's radius (the 
vertical solid line in Fig.~\ref{spacetime}). With our foliation, an
outgoing scalar wave is covered in only one slice.
 
\begin{figure}[t]
\centerline{\psfig{file=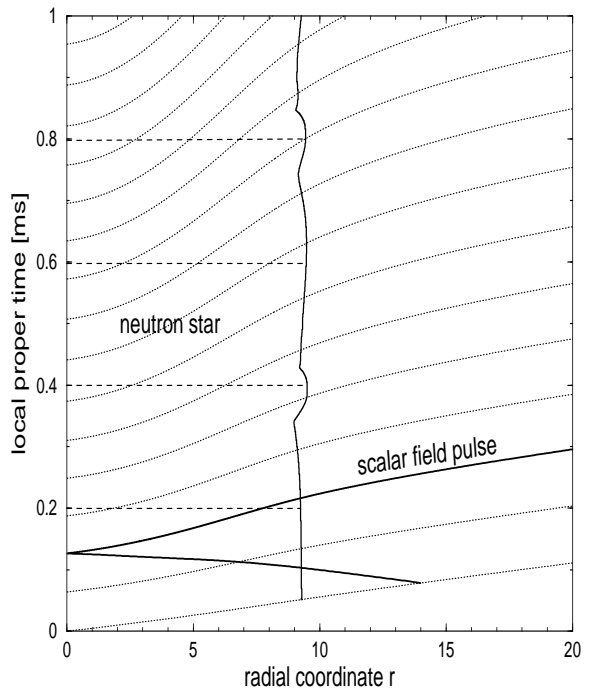,height=3.7in,width=3.0in,angle=-90}}
\caption{Spacetime diagram of the reflection of a Gaussian scalar field
pulse off a neutron star ($K=100, n=1$ and $\rho_c=1.5\times 10^{-3}$). 
The diagram focuses in the region close to the neutron star - scalar field 
interaction, but it was obtained from a global simulation of the spacetime. 
The dotted curves covering the whole diagram are outgoing light cones, 
which bend due to the spacetime curvature. The scalar field pulse, initially 
located at $r=14$, travels inwards, enters the neutron star and is 
reflected at the origin of the coordinate system (the solid line
corresponds to the maximum value of the scalar field). The interaction with 
the scalar field triggers the oscillation of the neutron star, which can 
be seen from the vertical solid line of varying location in the diagram, 
which indicates the radius of the star.}
\label{spacetime}
\end{figure}

Fig.~\ref{rhoc} displays the time evolution of the central
density of the different neutron stars in our setup. 
The solid lines correspond to the neutron star-scalar field 
system. Correspondingly, the dashed horizontal lines indicate
the evolutions of the equilibrium neutron star models {\it without}
the presence of the scalar field. As already mentioned these
evolutions are stable. They are 
characterized by the appearance of small-amplitude oscillations 
associated with the radial modes of pulsation of the star (which are
too small to be seen in the figure).  
On the other hand, all neutron star-scalar field models with initial 
central density below $2.8 \times 10^{-3}$ also oscillate 
around the stable equilibrium model. The oscillation frequencies
of the two least compact models correspond to the frequencies 
calculated in the linear regime, even though the amplitude of
the oscillations is now much larger due to the scalar field kick. 
This is no longer the case for the model with a central 
density of $\rho_c=2.8 \times 10^{-3}$. For this model, which is 
close to the threshold of black hole
formation, the amplitude of the oscillations is big enough to show
nonlinear effects, the oscillation frequency being much smaller than the
value obtained in the previous section. For the models with 
central densities of $\rho_c=2.9 \times 10^{-3}$ and 
$\rho_c=3.0 \times 10^{-3}$ the interaction with the scalar field is able 
to trigger their gravitational collapse to a black hole on a dynamical 
timescale. Unfortunately, due to numerical inaccuracies arising at the
end of the simulation we are not able to 
follow the collapse process once the event horizon is about to form 
(similar problems were reported in~\cite{LFJMP01} for the collapse 
of supermassive stars). However, convergence studies show clear evidence 
that these models collapse to black holes. Further evidence is given by 
the evolution of the neutron star radii, as shown in Fig.~\ref{radius}. 

To demonstrate the dynamic range of these evolutions,
we focus on the model with central density $\rho_{c} = 2.9 \times
10^{-3}$. Initially, the redshift factor $e^{2 H}$ between the 
center of the star and observers located at $r \to \infty$ is 2.1.
By the end of the simulation it has increased to a value of 59.5.
This high redshift factor explains the appearance of a
kink in the central density towards the end of our numerical 
evolution (see Fig.~\ref{rhoc}). 

We also note that global energy conservation
is very well fulfilled for these extreme hydrodynamic simulations.
The relative deviation from energy conservation according
to Eq.~(\ref{globcon}) when the evolution stops is of the order 
of $10^{-4}.$

\begin{figure}[t]
\centerline{\psfig{file=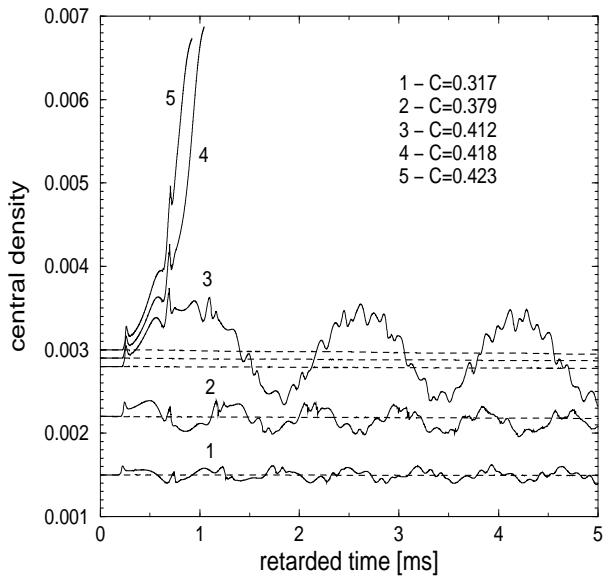,height=3.5in,width=3.5in}}
\caption{Central density of the neutron stars interacting with the
scalar field as a function of retarded time. The three less compact models, 
with $\rho_c \le 2.8\times 10^{-3}$, oscillate strongly around
their equilibrium value after interacting with the scalar field.
The other two more compact models collapse to a black hole instead
on a dynamical timescale. The dashed lines are taken from our evolutions 
of the equilibrium model without the presence of the scalar field.}
\label{rhoc}
\end{figure}

\begin{figure}[t]
\centerline{\psfig{file=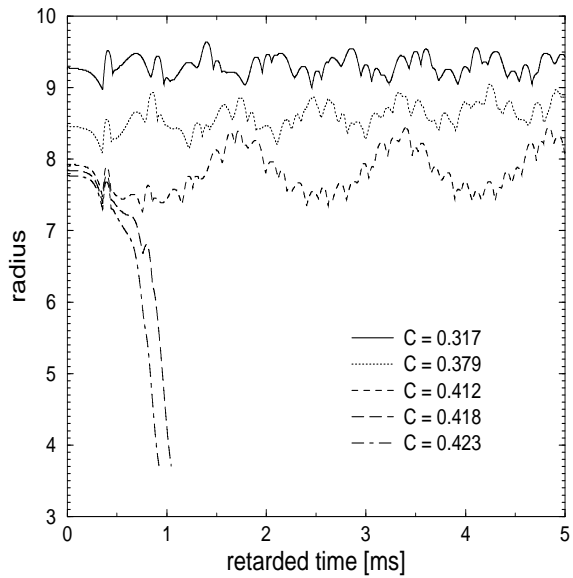,height=3.5in,width=3.5in}}
\caption{Time evolution of the radius of the different neutron stars 
interacting with the scalar field. The radius of two most compact models 
decreases dramatically, indicating that they undergo gravitational collapse 
to a black hole.}
\label{radius}
\end{figure}

By analyzing the energy transfer from the scalar field to the neutron
star during the interaction we find that it increases the more compact the
stellar models are. This behavior is shown in
Table~\ref{energy_transfer}. We remark that the initial mass of the
scalar field is not strictly identical for all cases considered, due to the
different underlying geometry on which the Gaussian pulse is
constructed.  We evaluate the total radiated mass in the scalar field
at a retarded time of $\tau=0.6$ms. The mass radiated away to
infinity after this time is negligible.

\begin{table}
\caption{
Energy transfer from the scalar field to the neutron star
during the scattering process. The entries are as follows: $\rho_{c}$ is
the central density of the neutron star, $M_0^{\Phi}$ is the initial mass
of the scalar field, $E_{\mbox{rad}}$ is the total radiated mass, and
$E_{\mbox{trans}}$ is the percentage of the energy transfered in the
interaction. We use units in which $G=c=M_{\odot}=1$.}
\medskip
\begin{tabular}{cccc}
$\rho_{c} \,\, (10^{-3})$ & $M_0^{\Phi}  \,\,(10^{-3})$ & 
$E_{\mbox{rad}} \,\, (10^{-3})$ & $E_{\mbox{trans}} \,\, (\%)$ \\
\hline
1.5  &  4.90  &  4.86  &  0.8
\\
2.2  &  4.80  &  4.72  &  1.7
\\
2.8  &  4.76  &  4.65  &  2.3
\\
2.9  &  4.75  &  4.63  &  2.5
\\
3.0  &  4.75  &  4.62  &  2.7
\\
\end{tabular}
\label{energy_transfer}
\end{table}

Next we analyze the behavior of the scalar field in these scattering
simulations. In Fig.~\ref{newszoom} we plot the (retarded) time evolution 
of the news, Eq.~(\ref{NNews}), for the whole sample of our neutron star 
models. The scalar field signal measured at null infinity can be neatly 
divided into three phases. The first phase, before the main pulse reflects 
off the origin (not shown in the figure), is dominated by an initial 
backscattering, the amplitude of the signal being small. 
The second phase, whose duration depends on the compactness of the 
neutron star~\cite{PTBS00}, is characterized by the reflection of 
the main scalar field pulse back and forth the origin
and the maximum of the neutron star curvature potential, which,
in turn, induces the appearance of quasi-normal oscillations on 
the scalar field. Most of the energy is radiated away in this 
period.  Once the pulse
has lost sufficient energy it enters a third phase, in which the
late time behavior of the signal is dominated by a power-law tail 
$N \propto t^{-\alpha}$, with $\alpha =3$, due to the reflection of the 
scalar field at 
the exterior Schwarzschild geometry~\cite{price,GPP94,GPP942,PTBS00}. 
Since the compactness of our models is well below the Buchdahl limit, 
$C=8/9$, the quasi-normal mode ringing phase does not last for an extended 
period of time. Therefore, after a few reflections trapped inside the curvature 
potential, the signal enters rapidly the power-law tail phase. From 
Fig.~\ref{newszoom} one can see that the more compact the neutron star, the 
larger the quasi-normal ringdown phase. 
We also point out that by going to more compact models, increasing the
central density of the neutron star beyond the maximum of the stability
curve (i.e., going into the unstable branch) and freezing the hydrodynamics
and metric evolution to avoid gravitational collapse, we are able
to find a much longer ringdown phase. Our results, obtained for fully 
self-gravitating, polytropic neutron star models, are in perfect agreement 
with previous findings by Pavlidou et al~\cite{PTBS00}, who used a more idealized
setup consisting of constant density, static neutron stars. 

\begin{figure}[t]
\centerline{\psfig{file=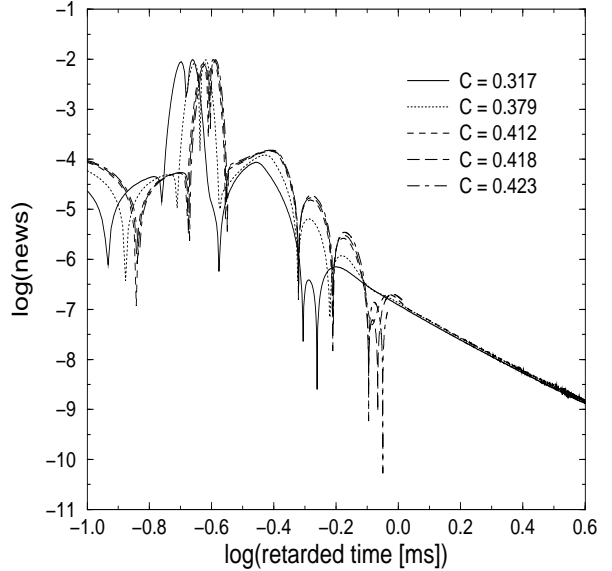,height=3.5in,width=3.5in}}
\caption{ Time evolution of the news function during the scattering
problem. The different lines correspond to the different models in our
sample of Table~\ref{nsmodels}, and are labeled in the plot with
respect to the compactness parameter.  The duration of the more
dynamic quasi-normal ringing phase strongly depends on the compactness
of the neutron star model, increasing as the compactness
increases. The late time behavior of the signal decays as an inverse
power-law.}
\label{newszoom}
\end{figure}

Since the study of the late time power-law tails requires sufficient
resolution, especially for large radii, we have used a different
radial coordinate for these simulations, $r = 30x/(1-x^{4})$. This
allowed us to resolve the power-law behavior, avoiding the
evolution from being dominated by numerical noise. By performing a 
linear regression study of the tails in the time interval 
$\log(\tau \left[ms\right]) \in
\left[0.3;0.7\right]$, we obtain the results summarized in
Table~\ref{power-law}. We find the correct power-law behavior of the
scalar field in our fully dynamical evolutions, as predicted by both,
linear analysis and by nonlinear numerical evolutions of scalar fields
in the exterior black hole geometry~\cite{GPP94,GPP942,GWS94}.  Note
that we measure the tails on the news, whereas the results of the
above references read off the quantity $g$ at future null infinity
$\cal{J}^{+}$. Both quantities are related by
\begin{equation}
N = e^{-2H} g_{,u}.
\end{equation}

\begin{table}
\centering
\begin{minipage}{4.5cm}
\caption{
Late time power-law behavior of the news $N \propto
t^{-\alpha}$ for the (stable) neutron star - scalar field scattering problem. 
The results agree with the predicted exact value $\alpha = 3$.}
\medskip
\begin{tabular}{cc}
$\rho_{c} \,\, (10^{-3})$ & $\alpha$ \\
\hline
1.5  & 3.06 
\\
2.2  & 3.05 
\\
2.8  & 3.05 
\\
\end{tabular}
\label{power-law}
\end{minipage}
\end{table}

\section{Summary}
\label{summary}

We have analyzed numerically the interaction of neutron stars and
scalar fields by means of nonlinear evolutions of the
Einstein-Klein-Gordon perfect fluid system in spherical symmetry. We
have built a sequence of stable, self-gravitating, $K=100$,
$n=1$ relativistic polytropes, increasing the central density from
$\rho_c=1.5\times 10^{-3}$ to $3.0\times 10^{-3}$ ($G=c=M_{\odot}=1$). 
Using a compactified spacetime foliation with outgoing null cones
we have studied the fate of the neutron stars when they are hit by 
a sufficiently strong scalar field packet, as well as the dynamics 
and energetics of the process. 

We have found that by choosing a strong (finite amplitude)
scalar field pulse with a mass of the order of $10^{-3} \ M_{\odot}$,
the neutron star is either forced to oscillate in its radial modes of
pulsations or to collapse to a black hole on a dynamical
timescale. The fate of the star depends on its central density and,
since we fix the polytropic equation of state, on its compactness. 
The energy transfered to the neutron star increases with the
compactness of the model. The radiative signals we have found 
in our fully nonlinear simulations consist of several quasi-normal
oscillations and a late time power-law tail, in agreement with the
results predicted by (linear) perturbation analysis of wave
propagation in an exterior Schwarzschild geometry~\cite{price}.

\section*{Acknowledgements}
It is a pleasure to thank Ewald M\"uller for useful discussions and
for a careful reading of the manuscript, and Johannes Ruoff by kindly
providing us the frequencies of the fundamental modes of the neutron
star models of Section~\ref{test:ns}, obtained with his perturbation
code. We further thank Nigel Bishop and Carsten Gundlach for helpful
comments on the manuscript. 
F.S. would like to thank the Relativity and Cosmology Group
of the University of Portsmouth, where part of this work was
done. This work was supported, in part, by the EU Programme
``Improving the Human Research and the Socio-Economic Knowledge Base"
(Research Training Network Contract HPRN-CT-2000-00137). P.P. also
acknowledges support from the Nuffield Foundation (award NAL/00405/G).

\end{document}